\documentstyle[11pt,a4,epsfig]{article}
\newcommand{\kslash}{k \hspace{-0.2cm} / \,}

\newcommand{\qslash}{q \hspace{-0.2cm} / \,}
\newcommand{\pqslash}{p_{1} \hspace{-0.34cm} / \,}
\newcommand{\psslash}{p_{2} \hspace{-0.34cm} / \,}

\begin{document}

\title{\bf Twist-3 Single Spin Asymmetries in Semi-Inclusive Deep-Inelastic
 Scattering}

\author{A.~Metz, M.~Schlegel
 \\[0.3cm]
{\it Institut f\"ur Theoretische Physik II,} \\
{\it Ruhr-Universit\"at Bochum, D-44780 Bochum, Germany}}

\date{\today}
\maketitle

\begin{abstract}
\noindent
The single spin asymmetries for a longitudinally polarized lepton beam 
or a longitudinally polarized nucleon target in semi-inclusive deep-inelastic 
scattering are twist-3 observables.
We study these asymmetries in a simple diquark spectator model of the
nucleon.
Analogous to the case of transverse target polarization, non-vanishing 
asymmetries are generated by gluon exchange between the struck quark and 
the target system. 
It is pointed out that the coupling of the virtual photon to the diquark 
is needed in order to preserve electromagnetic gauge invariance at the 
twist-3 level.  
The calculation indicates that previous analyses of these observables are
incomplete.
\end{abstract}

\section{Introduction}
The measurements of $A_{UL}$ (longitudinal target polarization) and
$A_{LU}$ (longitudinal lepton beam polarization) by the 
HERMES~\cite{HERMES_00,HERMES_01,HERMES_03} and CLAS~\cite{CLAS_03}
collaborations constitute the first clear evidence of non-vanishing
single spin asymmetries (SSA) in semi-inclusive deep-inelastic 
scattering (SIDIS) off the nucleon.
From a theoretical point of view, SSA in hard processes are very interesting 
because of their relation to time-reversal odd (T-odd) correlation
functions (parton distributions and fragmentation functions).

Already for more than a decade the existence of T-odd fragmentation 
functions is considered to be established~\cite{collins_93}.
In the meantime, explicit model calculations including final state interactions 
in the fragmentation process have provided non-vanishing results for such 
functions (see, e.g., Refs.~\cite{collins_94,bacchetta_01}).
On the other hand, it has been shown only recently that nonzero 
T-odd parton distributions are compatible with time-reversal invariance of 
the strong interaction~\cite{brodsky_02,collins_02} (see also 
Refs.~\cite{brodsky_02b,belitsky_03,boer_03} for related work).
In DIS, T-odd parton distributions arise due to the exchange of longitudinally
polarized gluons between the struck quark and the target system. 
This rescattering effect is encoded in the gauge link appearing in the 
definition of parton distributions. 
An alternative picture according to which T-odd parton distributions can
be generated without rescattering of the struck quark~\cite{anselmino_02} 
seems to be ruled out~\cite{pobylitsa_02}.

In particular, due to the recent developments~\cite{brodsky_02,collins_02}, 
the T-odd and transverse momentum dependent ($k_{\perp}$-dependent) 
functions $f_{1T}^{\perp}$ (Sivers function, describing the distribution of 
unpolarized quarks in a transversely polarized target)~\cite{sivers_90} 
and $h_1^{\perp}$ (distribution of transversely polarized quarks in an 
unpolarized target)~\cite{boer_98} may well exist.
From a practical point of view both distributions can be considered as
twist-2 functions, since they appear in observables at leading order
of a $1/Q$-expansion, where $Q$ denotes the large scale of the hard process.  
For instance, $f_{1T}^{\perp}$ enters the leading twist SSA $A_{UT}$ 
in SIDIS~\cite{boer_98}.

Despite of the recent progress in understanding the nature of T-odd effects,
a complete formalism (including subleading T-odd parton distributions) 
of T-odd twist-3 observables is still missing even at tree-level.
So far, such effects have only been treated on the fragmentation 
side~\cite{levelt_94,mulders_96}.
This point may also be quite important for the description of the twist-3 
asymmetries $A_{UL}$ and $A_{LU}$ in SIDIS.
With the exception of Refs.~\cite{yuan_03,gamberg_03}, all present 
analyses/calculations of these observables are based on~\cite{mulders_96}, 
i.e., they include only T-odd fragmentation functions 
(see e.g. Refs.~\cite{oganessyan_98,efremov_00,desanctis_00,ma_01,efremov_01,efremov_03}).
In this scenario one finds schematically $A_{LU} \propto e \, H_1^{\perp}$ and
$A_{UL} \propto h_L \, H_1^{\perp}$, with the twist-3 T-even parton distributions
$e$ and $h_L$, and the twist-2 T-odd Collins fragmentation function 
$H_1^{\perp}$~\cite{collins_93}.\footnote{In our calculation for $A_{UL}$ the target
is polarized along the direction of its momentum (and the direction of the virtual
photon). In experiments for the longitudinal target asymmetry, however, the 
polarization is along the direction of the incoming lepton. Both situations differ 
by a kinematical twist-3 term which is given by $A_{UT}$.}

Analogous to the treatment of $A_{UT}$ presented in Ref.~\cite{brodsky_02}, we 
compute $A_{LU}$ and $A_{UL}$ in the framework of a simple diquark spectator
model of the nucleon, in order to investigate whether T-odd parton distributions 
may be relevant in these cases.
The rescattering of the struck quark, which serves as the potential source of T-odd 
effects, is modelled by the exchange of an Abelian gauge boson.
Such a study has already been performed in Ref.~\cite{afanasev_03}.
However, in~\cite{afanasev_03} only $A_{LU}$ has been computed explicitly.
Moreover, the calculation of~\cite{afanasev_03} is not gauge invariant.
To preserve electromagnetic current conservation also the coupling of the 
virtual photon to the diquark has to be included.

Both $A_{LU}$ and $A_{UL}$ turn out to be nonzero indicating that, in a 
factorized picture, T-odd distributions have to be taken into account.
A first step in this direction has been made in~\cite{yuan_03}, where 
it has been demonstrated that the T-odd distribution $h_1^{\perp}$ appears in the 
description of $A_{LU}$ through a term $h_1^{\perp} \, E$, where $E$ is a 
twist-3 fragmentation function.
As will be discussed below, our calculation of $A_{LU}$, however, cannot 
be identified with such a term suggesting that the formula of~\cite{yuan_03} 
for the beam SSA is not yet complete.

\section{Tree diagrams}
In order to study SIDIS off a spin-$\frac{1}{2}$ particle (for definiteness
we think of a proton) in the framework of a spectator model we consider the 
process (compare also Ref.~\cite{brodsky_02})
\begin{equation} \label{e:process}
\gamma^{\ast}(q) + p(p,\lambda) \to q(p_1,\lambda') + s(p_2) \,.
\end{equation}
In full SIDIS, both the quark and the spectator in the final state fragment into 
hadrons, where we are interested in the situation that one of the hadrons from the 
quark fragmentation is detected at low transverse momentum.
However, for the study of possible T-odd effects related with parton distributions 
it is not necessary to include the fragmentation process in the calculation.
We use the model of~\cite{brodsky_02} with a scalar diquark spectator $s$.
In this model the proton has no electromagnetic charge, and a charge $e_1$
is assigned to the quark. 
The interaction between the proton, the quark and the spectator is described 
by a scalar vertex with the coupling constant $g$.

We treat the process~(\ref{e:process}) in the Breit frame of the virtual photon.
The proton has a large plus-momentum $Q/x$, where $x = x_{Bj} + {\cal O}(1/Q^2)$.
The quark carries the large minus-momentum $p_1^- \approx q^-$ and a soft 
transverse momentum $\vec{\Delta}_{\perp}$.
These requirements specify the kinematics:
\begin{eqnarray} \label{e:kin}
& & q = \Big( -Q, \, Q, \, \vec{0}_{\perp} \Big) , \qquad
    p = \bigg( \frac{Q}{x}, \, \frac{xM^2}{Q}, \, \vec{0}_{\perp} \bigg) ,
\\
& & p_1 = \bigg( \frac{\vec{\Delta}_{\perp}^2}{Q}, \, Q, \,
                 \vec{\Delta}_{\perp} \bigg) , \qquad
    p_2 = \bigg( \frac{Q (1-x)}{x}, \, 
                 \frac{x (\vec{\Delta}_{\perp}^2 + m_s^2)}{Q (1-x)}, \,
                -\vec{\Delta}_{\perp} \bigg) \,.
\nonumber
\end{eqnarray}
The expressions for $q$ and $p$ are exact, while for $p_1$ and $p_2$ just the
leading terms have been listed. 
In particular, sometimes the $1/Q^2$ corrections of $p_1^-$ and $p_2^+$ are 
needed which can be readily obtained from 4-momentum conservation.
To simplify the calculation we consider massless quarks.

\begin{figure}[t!]
\begin{center}
\includegraphics[width=13.0cm]{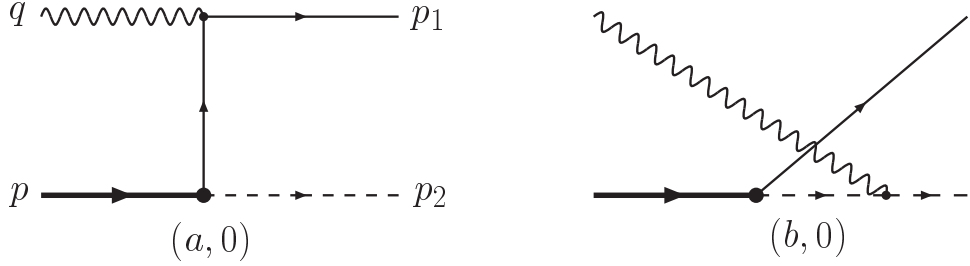}
\end{center}
\caption{Tree-level diagrams of the process in (\ref{e:process}).
 In order to preserve gauge invariance both diagrams have to be considered.
 \label{f:tree}}
\end{figure}

The tree-level diagrams of the process (\ref{e:process}) are shown in 
Fig.~\ref{f:tree}. 
Their currents, depending on the helicities of the proton and the quark, read
\begin{eqnarray}
J_{(a,0)}^{\mu}(\lambda,\lambda') & = &
 e_1 g \, \frac{1}{(p_1 - q)^2} \, \bar{u}(p_1,\lambda') \, \gamma^{\mu} \,
 (\pqslash - \qslash) \, u(p,\lambda) \,,
\\
J_{(b,0)}^{\mu}(\lambda,\lambda') & = &
 - e_1 g \, \frac{1}{(p_2 - q)^2 - m_s^2} \, (2p_2^{\mu} - q^{\mu}) \,
 \bar{u}(p_1,\lambda') \, u(p,\lambda) \,.
\end{eqnarray}
We have defined the current by means of the scattering amplitude
according to $T = \varepsilon_{\mu} J^{\mu}$, with $\varepsilon$ denoting the 
polarization vector of the virtual photon.
It is easy to check that current conservation holds for the sum of the 
two diagrams, i.e.,
\begin{equation}
q_{\mu} \Big( J_{(a,0)}^{\mu} + J_{(b,0)}^{\mu} \Big) = 0 \,.
\end{equation}

As long as one is just interested in leading twist observables, it is sufficient 
to consider diagram (a,0).
This is for instance the case in the calculation of the transverse SSA $A_{UT}$
in Ref.~\cite{brodsky_02}. 
The specific kinematics in Eq.~(\ref{e:kin}) is the reason for the suppression 
of diagram (b,0) relative to (a,0).
The propagator of the diquark in (b,0) behaves like $1/Q^2$, while there is
no large momentum flow through the quark propagator in (a,0).
Nevertheless, as we discuss in the following, for subleading twist observables
diagram (b,0) can no longer be neglected.

We compute the various components of the currents using the lightfront helicity 
spinors of~\cite{lepage_80}.
For $J_{(a,0)}^{\mu}$ one obtains
\begin{eqnarray} \label{e:a1}
J_{(a,0)}^{1}(\lambda,\lambda') & = &
 - e_1 g \, \frac{1-x}{\sqrt{x} \, |\vec{\Delta}_{\perp}|} \,
 \frac{Q}{\vec{\Delta}_{\perp}^2 + \tilde{m}^2} 
 \bigg[ Mx \Big(\Delta^1 - i\lambda \Delta^2 \Big) \delta_{\lambda,\lambda'}
       -\lambda \Big(\Delta^1 + i\lambda\Delta^2 \Big)^2 
       \delta_{\lambda,-\lambda'} \bigg] ,
\\  \label{e:a2}
J_{(a,0)}^{2}(\lambda,\lambda') & = &
 - e_1 g \, \frac{1-x}{\sqrt{x} \, |\vec{\Delta}_{\perp}|} \,
 \frac{Q}{\vec{\Delta}_{\perp}^2 + \tilde{m}^2} 
 \bigg[ Mx \Big(i\lambda \Delta^1 + \Delta^2 \Big) \delta_{\lambda,\lambda'}
       -i \Big(-i\lambda\Delta^1 + \Delta^2 \Big)^2 
       \delta_{\lambda,-\lambda'} \bigg] ,
\\  \label{e:aplus}
J_{(a,0)}^{+}(\lambda,\lambda') & = &
 - 2e_1 g \, \frac{1-x}{\sqrt{x}} \, 
 \frac{|\vec{\Delta}_{\perp}|}{\vec{\Delta}_{\perp}^2 + \tilde{m}^2} 
 \bigg[ Mx \, \delta_{\lambda,\lambda'}
       - \Big(\lambda \Delta^1 + i\Delta^2 \Big) 
       \delta_{\lambda,-\lambda'} \bigg] ,
\\  \label{e:aminus}
J_{(a,0)}^{-}(\lambda,\lambda') & = &
 - 2e_1 g \, \frac{1-x}{\sqrt{x}} \, 
 \frac{|\vec{\Delta}_{\perp}|}{\vec{\Delta}_{\perp}^2 + \tilde{m}^2} 
 \bigg[ Mx \, \delta_{\lambda,\lambda'}
       - \frac{x M^2}{\vec{\Delta}_{\perp}^2} 
      \bigg(1 - \frac{\vec{\Delta}_{\perp}^2 + m_s^2}{M^2 (1-x)} \bigg)
       \Big(\lambda \Delta^1 + i\Delta^2 \Big) 
       \delta_{\lambda,-\lambda'} \bigg] ,
\\
& &  \textrm {with} \quad 
 \tilde{m}^2 = x(1-x) \bigg(- M^2 + \frac{m_s^2}{1-x} \bigg) \,.
\nonumber
\end{eqnarray}
One observes here the well-known result that for DIS off a 
spin-$\frac{1}{2}$ particle the transverse current is dominating in the 
Breit frame.
For the second tree-graph we find
\begin{eqnarray} \label{e:bplus}
J_{(b,0)}^{+}(\lambda,\lambda') & = &
 - e_1 g \, \frac{2-x}{\sqrt{x} \, |\vec{\Delta}_{\perp}|} \, 
   \Big(\lambda \Delta^1 + i\Delta^2 \Big) \delta_{\lambda,-\lambda'} \,,
\\  \label{e:bminus}
J_{(b,0)}^{-}(\lambda,\lambda') & = &
 e_1 g \, \frac{\sqrt{x}}{|\vec{\Delta}_{\perp}|} \, 
   \Big(\lambda \Delta^1 + i\Delta^2 \Big) \delta_{\lambda,-\lambda'} \,.
\end{eqnarray}
\begin{figure}[t!]
\begin{center}
\includegraphics[width=13.0cm]{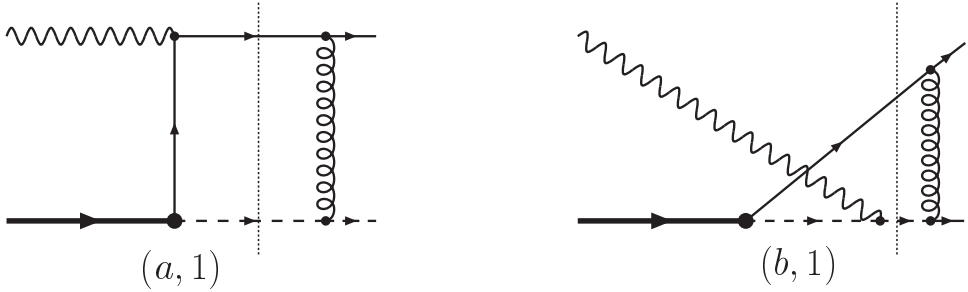}
\end{center}
\caption{One-loop diagrams for the imaginary part of the process in (\ref{e:process}).
 In each diagram the possible on-shell intermediate state is indicated
 by a thin dotted line. 
 The interaction between the quark and the spectator is modelled by the exchange
 of an Abelian gauge boson.
\label{f:loop}}
\end{figure}

The transverse components $J_{(b,0)}^i$ are proportional to $1/Q$ and, hence,
indeed are suppressed by a factor $1/Q^2$ compared to $J_{(a,0)}^i$ as 
expected.
Therefore, these terms are not relevant for the discussion of twist-3 
observables.
In contrast, the plus- and the minus-components for both diagrams are of the 
same order.
In this case, the suppression of (b,0) caused by the propagator of the diquark
is compensated by a factor $Q$ at the photon-diquark vertex and the fact 
that $\bar{u}(p_1,\lambda') \, u(p,\lambda) \propto Q$. 
Even though diagram (b,0) is not compatible with the parton model, since large
momentum transfers at the proton-quark-diquark vertex are allowed, a consistent 
calculation of twist-3 observables in the spectator model must contain 
this contribution.
We note that our results obey the gauge invariance constraint 
\begin{equation}
J_{(a,0)}^+ + J_{(b,0)}^+ = J_{(a,0)}^- + J_{(b,0)}^- \,.
\end{equation}
Including by hand a formfactor at the proton-quark-diquark vertex in order to suppress
large momentum transfers destroys the gauge invariance.

\section{One-loop diagrams}
To obtain nonzero SSA in a spectator model one has to go beyond the tree-level
approximation and take the rescattering of the quark into account.
For our purpose it is sufficient to model this effect by one-photon exchange.
Since the imaginary part of the one-loop amplitude is needed for the computation 
of SSA, just the two diagrams in Fig.~\ref{f:loop} have to be considered.
Self-energy and vertex correction diagrams are relevant for the real part
of the amplitude, but cannot acquire an imaginary part, because we are dealing
with either on-shell or even space-like (internal) lines.
In order to avoid infrared singularities at intermediate steps of the calculation
we assign a mass $\mu$ to the photon.
The final results for $A_{LU}$ and $A_{UL}$ must be infrared-finite which serves 
as a non-trivial check of the calculation.
The currents of the diagrams in Fig.~\ref{f:loop} are given by
\begin{eqnarray}
J_{(a,1)}^{\mu}(\lambda,\lambda') & = &
 i (e_1)^3 g \int \frac{d^4 k}{(2\pi)^4}
\\
& & \hspace{0.5cm} \mbox{} \times 
 \frac{\bar{u}(p_1,\lambda') \, (\pqslash + 2\psslash - \kslash) \,
 \kslash \, \gamma^{\mu} \, (\kslash - \qslash) \, u(p,\lambda)}
 {[k^2 + i\epsilon] [(k - q)^2 + i\epsilon]
  [(p + q - k)^2 - m_s^2 + i\epsilon] [(k - p_1)^2 - \mu^2 + i\epsilon]} \,,
\nonumber \\
J_{(b,1)}^{\mu}(\lambda,\lambda') & = &
 - i (e_1)^3 g \int \frac{d^4 k}{(2\pi)^4}
\\
& & \hspace{0.5cm} \mbox{} \times 
\frac{\bar{u}(p_1,\lambda') \, (\pqslash + 2\psslash - \kslash) \,
 \kslash \, (2p^{\mu} + q^{\mu} - 2k^{\mu}) \, u(p,\lambda)}
 {[k^2 + i\epsilon] [(p - k)^2 - m_s^2 + i\epsilon]
  [(p + q - k)^2 - m_s^2 + i\epsilon] [(k - p_1)^2 - \mu^2 + i\epsilon]} \,.
\nonumber
\end{eqnarray}
Applying Cutkosky-rules to calculate the imaginary part one can verify the
gauge invariance condition
\begin{equation}
q_{\mu} \Big( {\rm Im} \, J_{(a,1)}^{\mu} + {\rm Im} \, J_{(b,1)}^{\mu} \Big) = 0 \,. 
\end{equation}
It is obvious that the full current (including the real part) for the sum of 
both diagrams is not gauge invariant.

The calculation of the imaginary parts has been performed similar to the
study of T-odd fragmentation in Ref.~\cite{metz_02}.
We refrain from giving any details and just quote the final results.
For diagram (a,1) we find
\begin{eqnarray} \label{e:ima1}
{\rm Im} \, J_{(a,1)}^{1}(\lambda,\lambda') & = &
 - \frac{(e_1)^3 g}{8 \pi} \, 
   \frac{1-x}{\sqrt{x} \, |\vec{\Delta}_{\perp}|} \, Q \,
  \bigg[ \frac{\ln L}{\sqrt{\Lambda}}  
         Mx \Big(\Delta^1 - i\lambda \Delta^2 \Big) \delta_{\lambda,\lambda'}
\\
& & \hspace{1cm}  
 - \frac{1}{2 \vec{\Delta}_{\perp}^2} 
   \bigg( \Big(\vec{\Delta}_{\perp}^2 - \tilde{m}^2 + \mu^2 \Big)
          \frac{\ln L}{\sqrt{\Lambda}}  
        + \ln \frac{\tilde{m}^2}{\mu^2} \bigg)
        \lambda \Big(\Delta^1 + i\lambda\Delta^2 \Big)^2 
        \delta_{\lambda,-\lambda'} \bigg] ,
\nonumber \\ \label{e:ima2}
{\rm Im} \, J_{(a,1)}^{2}(\lambda,\lambda') & = &
 - \frac{(e_1)^3 g}{8 \pi} \, 
   \frac{1-x}{\sqrt{x} \, |\vec{\Delta}_{\perp}|} \, Q \,
  \bigg[ \frac{\ln L}{\sqrt{\Lambda}}  
         Mx \Big(i\lambda \Delta^1 + \Delta^2 \Big) \delta_{\lambda,\lambda'}
\\
& & \hspace{1cm}  
 - \frac{1}{2 \vec{\Delta}_{\perp}^2} 
   \bigg( \Big(\vec{\Delta}_{\perp}^2 - \tilde{m}^2 + \mu^2 \Big)
          \frac{\ln L}{\sqrt{\Lambda}}  
        + \ln \frac{\tilde{m}^2}{\mu^2} \bigg)
        i \Big(-i\lambda\Delta^1 + \Delta^2 \Big)^2 
        \delta_{\lambda,-\lambda'} \bigg] ,
\nonumber \\ \label{e:imaplus}
{\rm Im} \, J_{(a,1)}^{+}(\lambda,\lambda') & = &
 - \frac{(e_1)^3 g}{8 \pi} \, 
   \frac{1-x}{\sqrt{x} \, |\vec{\Delta}_{\perp}|} \, 
  \bigg[ Mx \bigg( \Big(\vec{\Delta}_{\perp}^2 - \tilde{m}^2 + \mu^2 \Big)
          \frac{\ln L}{\sqrt{\Lambda}}  
        + \ln \frac{\tilde{m}^2}{\mu^2} \bigg) \delta_{\lambda,\lambda'}
\\
& & \hspace{1cm}  
 + 2 \bigg( \tilde{m}^2 \frac{\ln L}{\sqrt{\Lambda}}
           - \ln \frac{Q^2 (1-x)}{\mu^2 x} + 1 \bigg)
        \Big(\lambda\Delta^1 + i\Delta^2 \Big) 
        \delta_{\lambda,-\lambda'} \bigg] ,
\nonumber \\
& & {\rm with} \quad 
 L = \frac{\vec{\Delta}_{\perp}^2 + \tilde{m}^2 + \mu^2 + \sqrt{\Lambda}}
          {\vec{\Delta}_{\perp}^2 + \tilde{m}^2 + \mu^2 - \sqrt{\Lambda}} \,,
 \quad
 \Lambda = \Big( \vec{\Delta}_{\perp}^2 + \tilde{m}^2 - \mu^2 \Big)^2
              + 4 \vec{\Delta}_{\perp}^2 \mu^2 \,.              
\nonumber
\end{eqnarray}
The plus-component of diagram (b,1) is given by
\begin{equation} \label{e:imbplus}
{\rm Im} \, J_{(b,1)}^{+}(\lambda,\lambda') = 
 - \frac{(e_1)^3 g}{8 \pi} \, \frac{1}{\sqrt{x} \, |\vec{\Delta}_{\perp}|} \, 
  \bigg( (2-x) \ln \frac{Q^2 (1-x)}{\mu^2 x} - 2(1-x) \bigg)
        \Big(\lambda\Delta^1 + i\Delta^2 \Big) 
        \delta_{\lambda,-\lambda'} \,.
\end{equation}
For the one-loop calculation we make use of gauge invariance to eliminate
the minus-component of the current.
The $Q$-behaviour of the one-loop expressions corresponds to the one of the
tree-graphs.
Note also that the plus-component of the currents for both diagrams contains 
a $\ln Q^2$-term, which is not compatible with the parton model.
From our results for the transverse currents in 
Eqs.~(\ref{e:ima1},\ref{e:ima2}) we were able to reproduce the 
transverse target SSA computed in Ref.~\cite{brodsky_02} 
(up to an overall sign).

\section{Spin asymmetries}
Eventually, we proceed to the calculation of $A_{LU}$ and $A_{UL}$.
The full cross section in DIS (including the leptons) in the one-photon exchange 
approximation can be expressed in the standard form
\begin{equation}
\sigma \propto L_{\mu\nu} W^{\mu\nu} \,, 
\end{equation}
with the lepton tensor
\begin{equation} \label{e:ltensor}
L^{\mu\nu} = 2 \Big( l^{\mu} l'^{\nu} + l^{\nu} l'^{\mu} 
           - \frac{Q^2}{2} g^{\mu\nu} 
           + i \lambda_e \varepsilon^{\mu\nu\rho\sigma}q_{\rho}l_{\sigma} \Big) \,.
\end{equation}
In Eq.~(\ref{e:ltensor}), the 4-momentum of the incoming (outgoing) lepton is denoted
by $l$ ($l'$), and $Q^2 = -(l-l')^2$.
The hadron tensor is obtained from the above currents according to
\begin{equation}
W^{\mu\nu} = \Big( J^{\mu} \Big)^{\dagger} J^{\nu} \,.
\end{equation}
Now, we exploit gauge invariance of both the lepton and hadron tensor, take $q$ from
Eq.~(\ref{e:kin}), choose the lepton momenta to be in the $xz-$plane, and ignore
a contribution of $W^{++}$ which for our calculation is at least suppressed by a 
factor $1/Q^2$ relative to the leading term in the cross section.
This allows us to write the cross section as
\begin{eqnarray} \label{e:sigma}
\sigma & \propto & \frac{4Q^2}{y^2} \bigg[ - (2-y) \sqrt{1-y} \, W_S^{+1}
       + \Big( 1 - y + \frac{y^2}{4} \Big) W_S^{11} + \frac{y^2}{4} \, W_S^{22}
\\
& & \hspace{1cm} + i \lambda_e \bigg( y \sqrt{1-y} \, W_A^{+2}
       - y \Big( 1 - \frac{y}{2} \Big) W_A^{12} \bigg) \bigg] ,
\nonumber     
\end{eqnarray}
where $W_S^{\mu\nu} = (W^{\mu\nu} + W^{\nu\mu})/2$ and 
$W_A^{\mu\nu} = (W^{\mu\nu} - W^{\nu\mu})/2$ represent the symmetric and the 
antisymmetric part of the hadronic tensor respectively.
We also used the standard definition $y = p \cdot q / p \cdot l$.
The first line in Eq.~(\ref{e:sigma}) is relevant for the target spin asymmetry, 
and the second one for the beam asymmetry.
Actually, it turns out that the purely transverse components of the hadronic 
tensor $(W_S^{11}, \; W_S^{22}, \; W_A^{12})$ don't contribute to the 
spin-asymmetries at twist-3 level.

In order to specify the asymmetries we define
\begin{eqnarray} \label{e:w_uu}
W_{UU}^{11} & = & \frac{1}{2} \sum_{\lambda,\lambda'}
 \Big( J^1 (\lambda,\lambda') \Big)^{\dagger} J^1 (\lambda,\lambda') \,,
\\  \label{e:w_alu}
W_{A,LU}^{+2} & = & \frac{1}{4} \sum_{\lambda,\lambda'}
 \bigg[ \Big( J^+ (\lambda,\lambda') \Big)^{\dagger} J^2 (\lambda,\lambda')
      - \Big( J^2 (\lambda,\lambda') \Big)^{\dagger} J^+ (\lambda,\lambda')
 \bigg] \,,
\\  \label{e:w_sul}
W_{S,UL}^{+1} & = & \frac{1}{4} \sum_{\lambda'}
 \bigg[ \bigg( \Big( J^+ (\uparrow,\lambda') \Big)^{\dagger} 
                     J^1 (\uparrow,\lambda')
      - \Big( J^+ (\downarrow,\lambda') \Big)^{\dagger} 
              J^1 (\downarrow,\lambda') \bigg)      
\\ 
& & \hspace{1.0cm} + \bigg( \Big( J^1 (\uparrow,\lambda') \Big)^{\dagger} 
                                  J^+ (\uparrow,\lambda')
      - \Big( J^1 (\downarrow,\lambda') \Big)^{\dagger} 
              J^+ (\downarrow,\lambda') \bigg)      
\bigg] \,,
\nonumber
\end{eqnarray}
where polarization ``$\uparrow$'' in (\ref{e:w_sul}) means polarization
along the positive $z$-axis, i.e., along the direction of the target
momentum.
The element $W_{UU}^{11}$ is given by the tree-level result for diagram 
(a,0) in Eq.~(\ref{e:a1}).
Nonzero contributions to $W_{A,LU}^{+2}$ and $W_{S,UL}^{+1}$ are generated
by interference of the tree-level amplitude with the imaginary part of the 
one-loop amplitude.
While the transverse currents in~(\ref{e:w_alu}) and~(\ref{e:w_sul}) are obtained 
from diagrams (a,0) and (a,1), all four diagrams contribute to the plus-component 
of the current. 
The final results for the asymmetries read\footnote{As a reference we 
mention that the transverse target spin asymmetry of~\cite{brodsky_02} is given by
$A_{UT} = W_{UT}^{11} / W_{UU}^{11}$, with $W_{UT}^{11} = W_{S,UT}^{11}$ defined
analogous to Eq.~(\ref{e:w_sul}).}
\begin{eqnarray} \label{e:lu_res}
A_{LU} & = & \frac{i W_{A,LU}^{+2}}{W_{UU}^{11}}
 = \frac{(e_1)^2}{4\pi} \frac{\vec{\Delta}_{\perp}^2 + \tilde{m}^2}
                                   {M^2 x^2 + \vec{\Delta}_{\perp}^2} 
         \frac{\Delta^2}{Q} 
   \bigg[ \frac{1}{\vec{\Delta}_{\perp}^2} 
                   \bigg( -M^2 x^2 - \tilde{m}^2 \frac{2-x}{2(1-x)} \bigg)
                   \ln \frac{\vec{\Delta}_{\perp}^2 + \tilde{m}^2}{\tilde{m}^2}        
\\
& & \hspace{5.5cm}        
          - \frac{x}{2(1-x)} 
          \ln \frac{Q^2 (1-x)}{(\vec{\Delta}_{\perp}^2 + \tilde{m}^2)x} 
   \bigg] ,
\nonumber \\  \label{e:ul_res}
A_{UL} & = & \frac{W_{S,UL}^{+1}}{W_{UU}^{11}}
 = \frac{(e_1)^2}{4\pi} \frac{\vec{\Delta}_{\perp}^2 + \tilde{m}^2}
                                   {M^2 x^2 + \vec{\Delta}_{\perp}^2} 
         \frac{\Delta^2}{Q} 
   \bigg[ \frac{1}{\vec{\Delta}_{\perp}^2} 
                   \bigg( M^2 x^2 - \tilde{m}^2 \frac{2-x}{2(1-x)} \bigg)
                   \ln \frac{\vec{\Delta}_{\perp}^2 + \tilde{m}^2}{\tilde{m}^2}        
\\
& & \hspace{5.5cm}        
          - \frac{x}{2(1-x)} 
          \ln \frac{Q^2 (1-x)}{(\vec{\Delta}_{\perp}^2 + \tilde{m}^2)x} 
   \bigg] .
\nonumber
\end{eqnarray}
We would like to add some remarks:
\begin{itemize}
\item An explicit nonzero result for $A_{LU}$ in the framework of the diquark
 spectator model has already been obtained in Ref.~\cite{afanasev_03}.
 Our calculation shows that both $A_{LU}$ and $A_{UL}$ remain finite once all 
 diagrams required by electromagnetic gauge invariance are taken into 
 consideration.
\item We believe that the effect which generates $A_{LU}$ in our calculation 
 is not related to a term proportional to $h_1^{\perp} \, E$ discussed 
 in Ref.~\cite{yuan_03}. 
 While $h_1^{\perp}$ is chirally odd, we have summed over the polarizations of 
 the outgoing quark.
\item Since the asymmetries are proportional to 
 $\Delta^2 = |\vec{\Delta}_{\perp}| \sin \phi_q$ we expect in full
 SIDIS an effect proportional to $\sin \phi_h$, where $\phi_h$ is the azimuthal 
 angle of the produced hadron.
 The mechanisms which have been discussed so far in the literature in connection 
 with $A_{LU}$ and $A_{UL}$~\cite{levelt_94,mulders_96,yuan_03} show the same 
 $\sin \phi_h$-behaviour.
 In addition, the different contributions to the asymmetries have the same 
 $y$-dependence ($y\sqrt{1-y}$ for $A_{LU}$ and 
 $(2-y)\sqrt{1-y}$ for $A_{UL}$, see Eq.~(\ref{e:sigma})).
\item In the final results for the asymmetries we have performed the limit 
 $\mu \to 0$ without encountering a divergence.
 We agree with the observation made in Ref.~\cite{afanasev_03}, that the contribution
 from diagrams (a,0) and (a,1) to $A_{LU}$ is separately infrared-finite.
 This behaviour, which holds for $A_{UL}$ as well, seems to be accidental.
\end{itemize}

\section{Summary and conclusions}
In summary, we have calculated the twist-3 single spin asymmetries $A_{LU}$ and
$A_{UL}$ for semi-inclusive DIS off a nucleon target in the framework of a 
simple diquark spectator model.
Our study completes the previous work in Ref.~\cite{afanasev_03}, where only
$A_{LU}$ has been computed explicitly.
Moreover, the treatment in~\cite{afanasev_03} is lacking electromagnetic gauge 
invariance.

Both $A_{LU}$ and $A_{UL}$ turn out to be nonzero.
Although the spectator model calculation contains contributions which are not
compatible with the parton model, the non-vanishing results indicate that T-odd 
distributions have to be included in a factorized description of the asymmetries.
So far this has only been done partly in the literature.
In fact, we have argued that apparently none of the present analyses/calculations 
of $A_{LU}$ and $A_{UL}$ within the parton model is complete.
Mainly for two reasons we feel confident to make such a speculation:
first, within our calculation non-zero asymmetries arise already from the 
diagram (a,1), whose kinematics is compatible with the parton model. 
Second, there is no reason why the asymmetries should not contain higher
order T-odd distribution functions.
The status of the parton model formulae for $A_{LU}$ and $A_{UL}$ needs to be 
clarified before any definite conclusion can be extracted from the data.
\\[0.6cm]
\noindent
{\bf Note added:} After this work has been completed, a revised parton model
analysis for $A_{LU}$ and $A_{UL}$ appeared~\cite{bacchetta_04}.
The analysis confirms our suspicion that both asymmetries should contain an 
additional term with a twist-3 T-odd distribution function, which have not been
taken into account in the literature before.
\\[0.6cm]
\noindent
{\bf Acknowledgements:}
We are grateful to J.C.~Collins and N.~Kivel for discussions.
The work has been partly supported by the Sofia Kovalevskaya Programme of the
Alexander von Humboldt Foundation, the DFG and the COSY-Juelich project.

\end{document}